\documentclass[english,12pt,aps,prd,a4paper,preprintnumbers,floatfix,nofootinbib,showpacs,superscriptaddress, notitlepage]{revtex4-1} 
 \pdfoutput=1
\usepackage[usenames,dvipsnames]{color}  %%%%%%% usenames,dvipsnames required to get BrickRed
\usepackage{graphicx}
\usepackage{caption}
\usepackage{subcaption}
\usepackage[export]{adjustbox} % loads also graphicx
\usepackage{amsmath}
\usepackage{amssymb}
\usepackage{amsfonts}
\usepackage{bbold}
\usepackage[colorlinks=true,citecolor=darkred,urlcolor=darkred, pdfborder={0 0 0}]{hyperref}
\usepackage[normalem]{ulem}
\usepackage{braket}

\makeatletter
\def\p@subsection{}
\makeatother

\definecolor{darkred}{rgb}{0.6,0,0}

\definecolor{linkcolor}{rgb}{0,0,0.5}

\def\gsim{\raise0.3ex\hbox{$\;>$\kern-0.75em\raise-1.1ex\hbox{$\sim\;$}}}
\def\lsim{\raise0.3ex\hbox{$\;<$\kern-0.75em\raise-1.1ex\hbox{$\sim\;$}}}

\def\beqn#1{\begin{equation}\label{#1}}
\def\eeqn{\end{equation}}

\def\beqa#1{\begin{eqnarray}\label{#1}}
\def\eeqa{\end{eqnarray}}

\def\0nbb {$0\nu\beta\beta$ }

\def\Z2{$\mathcal{Z_2}$}

\newcommand {\ignore}[1]{}

\def\321{$\mathrm{SU(3) \otimes SU(2) \otimes U(1)}$ }

\newcommand{\AddrAHEP}{
	AHEP Group, Institut de F\'{i}sica Corpuscular,
	CSIC/Universitat de Val\`{e}ncia, Parc Cient\'ific de Paterna.\\
	C/ Catedr\'atico Jos\'e Beltr\'an, 2 E-46980 Paterna (Valencia), Spain}
\newcommand{\AddrHBNI}{
	Homi Bhabha National Institute, BARC Training School Complex, Anushakti Nagar, Mumbai 400094, India }

\begin{document}

\bibliographystyle{unsrt} 

\title{ALP-portal majorana dark matter}
\author{Shivam Gola}
\email{shivamg@imsc.res.in}
\affiliation{The Institute of Mathematical Sciences,
C.I.T Campus, Taramani, Chennai 600 113, India}
\affiliation{\AddrHBNI}
\author{Sanjoy Mandal}
\email{smandal@ific.uv.es}
\affiliation{\AddrAHEP}
\author{Nita Sinha}
\email{nita@imsc.res.in}
\affiliation{The Institute of Mathematical Sciences,
C.I.T Campus, Taramani, Chennai 600 113, India}
\affiliation{\AddrHBNI}

\begin{abstract}
\vspace{0.5cm}
Axion like particles(ALPs) and right handed neutrinos~(RHNs) are two well-motivated dark matter(DM) candidates. However, these two particles have a completely different origin. Axion was proposed to solve the Strong CP problem, whereas RHNs were introduced to explain light neutrino masses through seesaw mechanisms. We study the case of ALP portal RHN DM~(Majorana DM) taking into account existing constraints on ALPs. We consider the leading effective operators mediating interactions between the ALP and SM particles and three RHNs to generate light neutrino masses through type-I seesaw. Further, ALP-RHN neutrino coupling is introduced to generalize the model which is restricted by the relic density and indirect detection constraint.
\end{abstract}
%%%%%%%%%%%%%%%%%%%%%%%%%%%%%%%%%
\maketitle
%%%%%%%%%%%%%%%%%%%%%%%%%%%%%%%%%%%%
\section{Introduction}
%%%%%%%%%%%%%%%%%%%%%%%%%%%%%%%%%%%%%%%%%%%
Dark matter is one of the most important issues of modern particle physics and cosmology. Although a wide variety of experiments ranging from sub-galactic scale to a large cluster of galaxies have accumulated data in support of DM's existence~\cite{Bartelmann:1999yn,Clowe:2003tk,Harvey:2015hha,Hinshaw:2012aka,Ade:2015xua}, its microscopic properties still remain unknown. Several DM candidates have been proposed and searched for, however, no completely satisfactory DM candidate has been found so far. Among a lot of possibilities, the axion and sterile neutrinos can be regarded as a leading candidates for DM. These particles arise in well-motivated extensions of the Standard Model and have very rich phenomenology.

%%%%%%%%%%%%%%%%%%%%%%%%%%%%%%%%%%%%%%%%%%%%%%%%%%%%%%%%%%%%%%%%
It is known from the cosmological observations that the sum of the active neutrino masses $\sum m_{\nu}\leq 0.12$ eV~\cite{Aghanim:2018eyx,Lattanzi:2017ubx} and contribution to the relic density $\leq4.5\times10^{-3}$, which is too low to explain the DM abundances today, hence there is no explanation for DM within the SM. However, a new heavier neutrino field could explain DM and it is naturally required to explain the masses of active neutrinos as inferred from the oscillation experiments~\cite{deSalas:2020pgw}. The discovery of neutrino oscillations confirming the existence of at least two non-vanishing neutrino mass-squared differences necessitate physics beyond the Standard Model~(BSM). In principle neutrino mass could be simply generated by addition of right-handed neutrinos~(RHNs) to the SM particle content. These RHNs interact with SM fields via mixing with active neutrinos. Since RHNs are SM singlet, they allow Majorana mass term along with usual Dirac mass term. This is known as type-I seesaw mechanism~\cite{Minkowski:1977sc,Schechter:1980gr,Mohapatra:1979ia,Schechter:1981cv}. Mass of these RHNs could range from eV to GUT scale depending on the models~\cite{Dorsner:2006fx,Bajc:2007zf,deGouvea:2006gz,deGouvea:2007hks}. RHNs can also play the role of warm dark matter(WDM), which is singlet under the SM gauge symmetry and has tiny mixing with the SM neutrinos leading to a long lifetime~\cite{Abada:2014vea,Borah_2016,Das:2019kmn}. Also, KeV scale RHNs have been studied as a viable DM candidate~\cite{Merle:2017jfn,Adhikari:2016bei,Abada:2014zra}. In this work we have instead focused on the prospects of having GeV scale RHNs as a Weakly interacting massive particle~(WIMP) DM candidate.

%%%%%%%%%%%%%%%%%%%%%%%%%%%%%%%%%%%%%%%%%%%%%%%%%%%%%%%%%%%%%%%%%%%%%%
Axion~\cite{Peccei:1996ax} was postulated in the Peccei-Quinn (PQ) mechanism to solve the strong CP problem~\cite{Peccei:2006as,Kim:2008hd,Hook:2018dlk,Lombardo:2020bvn} of quantum chromodynamics (QCD). This axion can be identified as a (pseudo) Nambu-Goldstone boson associated with the spontaneous breaking of the $U(1)_{\text{PQ}}$ global symmetry~\cite{Weinberg:1977ma,Wilczek:1977pj,Berezhiani:1989fp}. This QCD axion gets a tiny mass from the explicit breaking of this global symmetry due to QCD anomaly. Astrophysical and experimental searches have not favoured the PQ model. To resolve issues with PQ model, other popular solutions like KSVZ~\cite{Kim:1979if,Shifman:1979if}, DFSZ~\cite{Dine:1981rt} etc. invoking axion were also proposed and studied afterwards. The magnitude of the couplings of axions to ordinary matter is inversely proportional to the axion decay constant $f_a$ which is associated with the $U(1)_{\text{PQ}}$ symmetry breaking scale. Hence, the couplings are highly suppressed if $f_a$ is sufficiently large and this features make the axion suitable to be a DM candidate. Many BSM extensions which features spontaneously broken global $U(1)$ symmetry predict massless Nambu-Goldstone bosons whose couplings are not constrained unlike the original QCD axion. These kind of particles are known as axion like particles~(ALPs). Mass of these ALPs are not related to its symmetry breaking scale unlike PQ axion. In general they are not supposed to solve the strong CP problem, but with the introduction of planck scale operators they could solve the strong CP problem~\cite{Hook:2019qoh,Kelly:2020dda}. Here we will consider the most general $SU(2)_L\otimes U(1)_Y$ invariant formulation of ALP interactions developed in Refs.~\cite{Georgi:1986df,Brivio:2017ije,Salvio:2013iaa}. This generic effective ALPs Lagrangian allows ALPs coupling with all the SM gauge bosons as well as with all the SM fermions. In addition to this effective ALP Lagrangian, we introduce three RHNs which can generate light neutrino masses through type-I seesaw. We invoke a $\mathbb{Z}_2$ symmetry under which all SM and BSM particles are even except the lightest RHN. Further we introduced the RHN-ALP coupling and show that the lightest RHN~(Singlet majorana fermion) which is odd under $\mathbb{Z}_2$ can play the role of DM candidate. Note that the phenomenology of this ALP-mediated DM~\cite{Hochberg:2018rjs} will be similar to pseudoscalar-portal DM. Also DM interacting via the exchange of a light ALP can induce observable signals in indirect detection experiments while evading the strong bounds from direct DM searches. Note that ALPs with mass $M_a\sim\mathcal{O}(\text{GeV})$ may also show up at colliders~\cite{Mimasu:2014nea,Jaeckel:2015jla,Alves:2016koo} or in rare meson decays~\cite{Dolan:2014ska,Izaguirre:2016dfi,CHOI1986145}.\\
%%%%%%%%%%%%%%%%%%%%%%%%%%%%%%%%%%%%%%%%%%%%%%%%%%%%%%%%%

Rest of the manuscript is organized as follows. In Sec.~\ref{sec:Model} we introduced our model, detailing the new interactions present. In Sec.~\ref{sec:Constraints} we summaries the existing constraints on ALP parameter space coming from various observables and collider searches. In Sec.~\ref{sec:Dark matter} we have explored and discussed the feasible parameter space coming from DM analyses such as relic density, direct and indirect detection. Finally, we give our conclusions in Sec.~\ref{sec:Conclusion}.
%%%%%%%%%%%%%%%%%%%%%%%%%%%%%%%%%%%%%%%%%%%%%%%%%%%%%%%%%%%%%%%%
%%%%%%%%%%%%%%%%%%%%%%%%%%%%%%%%%%%%%%%%%%%%%%%%%%%%%%%%%%%%%%%%%%%%%%%%
\section{Model}
\label{sec:Model}
%%%%%%%%%%%%%%%%%%%%%%%%%%%%%%%%%%%%%%%%%%%%%%%%%%%%%%%%%%%%%%%%%%
The model that we consider is the minimal combination of type-I seesaw and effective ALP interaction with additional $\mathbb{Z}_2$ symmetry apart from the SM gauge symmetry~\cite{Salvio:2021puw,Salvio:2015cja}. The matter content of the model is shown in Table.~\ref{tab:quantun-numbers}. Let's first briefly discuss the features of a generic ALP Lagrangian. We extend the SM particle content by adding an additional ALP which is a singlet under SM charges and is a pseudo Nambu-Goldstone boson of a spontaneously broken symmetry at some energy which is higher than the electroweak scale $v$. In effective theory the operators will be weighted by powers of $a/f_a$, where $f_a$ is the scale associated to the physics of the ALP, $a$. Effective linear Lagrangian with one ALP has been already discussed in great detail in Ref.~\cite{Georgi:1986df,Brivio:2017ije}. For linear EWSB realizations the most general linear bosonic Lagrangian involving $a$ is given by,
\begin{table}[!t]
\centering
\begin{tabular}{|c||c|c|c|c|c|c||c|c||c|c|}
\hline
        & \multicolumn{6}{|c||}{Standard Model} &  \multicolumn{2}{|c||}{New Fermions}  & \multicolumn{1}{|c|}{New Scalar}  \\
        \cline{2-10}
        & \hspace{0.2cm} $\ell_L$  \hspace{0.2cm} &  \hspace{0.2cm} $e_R$ \hspace{0.2cm} & \hspace{0.2cm} $q_L$ \hspace{0.2cm} & \hspace{0.2cm} $u_R$ \hspace{0.2cm} & \hspace{0.2cm} $d_R$  \hspace{0.2cm} &  \hspace{0.2cm}$H$ \hspace{0.2cm}  &  \hspace{0.2cm} $N_1$ \hspace{0.2cm} &  \hspace{0.1cm} $N_{2,3}$  \hspace{0.1cm}  &  \hspace{0.01cm}$a$   \hspace{0.01cm}\\
\hline     
% types & 3 & 3 & 1 & 1 & 1 & 1  \\                            
$SU(2)_L$ &  2    &  1  & 2  & 1 & 1  &    2    &     1    &  1   &    1       \\
\hline
$U(1)_Y$  & -1/2    &  -1 & 1/6  & 2/3  & -1/3    &    1/2    &     0    &  0   &   0    \\
\hline
$\mathbb{Z}_2$   &  $+$  &  $+$  & $+$ &  $+$ & $+$  &  $+$  &  $-$   & $+$  &  $+$     \\
\hline
\end{tabular}
\caption{Matter content and charge assignment of the considered model.}
\label{tab:quantun-numbers}
\end{table}
%%%%%%%%%%%%%%%%%%%%%%%%%%%%%%%%%%%%%%%%%%%%%%%%%%%%%%%%%%%%%%%%%%%%
\begin{align}
\mathcal{L}  =  \mathcal{L}_{\text{SM}} +  \mathcal{L}_{\text{ALP}},
\end{align}
%%%%%%%%%%%%%%%%%%%%%%%%%%%%%%%%%%%%%%%%%%%%%%%%%%%%%%%%%%%%%
where the leading order effective Lagrangian $\mathcal{L}_{\text{SM}}$ is same as the SM one and with
%%%%%%%%%%%%%%%%%%%%%%%%%%%%%%%%%%%%%%%%%%%%%%%%%%%%%%%%%%%%
\begin{align}
\mathcal{L}_{\text{ALP}} \ &= \ \frac{1}{2}\partial_{\mu}a \partial^{\mu}a \ - \ \frac{1}{2} M_a^2 a^2  -  \frac{C_{\tilde{G}}}{f_a} \ a G_{a\mu\nu}\tilde{G}^{\mu\nu a} -  \frac{C_{\tilde{B}}}{f_a} \ a B_{\mu\nu}\tilde{B}^{\mu\nu} - \frac{C_{\tilde{W}}}{f_a} \ a W_{a\mu\nu}\tilde{W}^{a\mu\nu} \nonumber \\
& + i C_{a\Phi}\times \big[(\bar{Q}_L Y_U \tilde{\Phi}u_R - \bar{Q}_L Y_D \Phi d_R - \bar{L}_L Y_E \Phi e_R )\frac{a}{f_a} + h.c. \big] 
\end{align}
%%%%%%%%%%%%%%%%%%%%%%%%%%%%%%%%%%%%%%%%%%%%%%%%%%%%%%
Here $\Phi$ and $a$ are the Higgs and ALP fields respectively. $C_{i}$ where $i=G,B,W,a\Phi$ are the corresponding Wilson coefficients for ALP-gauge boson and ALP-matter interactions. Parameters $M_a$ and $f_a$ are the ALP mass and energy scale associated to ALP physics. $Y_D$, $Y_U$ and $Y_E$ are $3\times 3$ matrices in flavour space which stands for down-type quarks, up-type quarks and charged leptons, respectively. We see that ALP Lagrangian has a very generic form than that of the QCD-axion, where the mass of ALP is not restricted by the new physics scale. The remaining fields and parameters are the same as in SM. References \cite{alves2020probing, Brivio:2017ije, Atre:2009rg} have discussed the phenomenology of the various pieces of the model. Now let's introduce the type-I seesaw Lagrangian with three additional SM singlet RHNs:
%%%%%%%%%%%%%%%%%%%%%%%%%%%%%%%%%%%%%%%%%%%%%%%%
\begin{align}
\mathcal{L}_{\text{RHN}} \ =  \ i\sum^3_{i=1}\bar{N}_i\gamma^{\mu}\partial_{\mu} N_i \ - \sum^3_{j=2} Y_{\alpha j} \bar{L}_\alpha \tilde{\Phi} N_j  \ - \ \sum^3_{i,j=2}M_{ij} \bar{N}^c_i N_j \ - \ M_{N_1} \bar{N}^c_1 N_1 \ + \ \text{h.c.} 
\end{align}
%%%%%%%%%%%%%%%%%%%%%%%%%%%%%%%%%%%%%%%%%%%%%%
where $N_i$ are the SM singlet RHNs, $Y_{\alpha j}$ is the Dirac Yukawa coupling and $M_{ij}$ is the Majorana mass term. As, $N_1$ is odd under $\mathbb{Z}_2$ symmetry, Dirac type of Yukawa interaction is forbidden for it unlike that for $N_{2,3}$. As a result of this, one of the light neutrinos will remain massless and we have enough parameters to describe the neutrino oscillations data. In addition to this, $\mathbb{Z}_2$ symmetry stabilizes the $N_1$ and it can play the role of DM candidate if one allows the following ALP-RHN interaction,
%%%%%%%%%%%%%%%%%%%%%%%%%%%%%%%%%%%%%%%%%%%%%%%%%%%
\begin{align}
\mathcal{L}_{\text{ALP-RHN}} \  =  - \sum^3_{i=1} \frac{C_{aN_{i}}}{f_{a}} (\bar{N_{i}}\gamma^{\mu}\gamma^{5}N_i) \partial_{\mu}a,
\end{align}
%%%%%%%%%%%%%%%%%%%%%%%%%%%%%%%%%%%%%%%%%%%%%%%%%%%%%%%%%%%%%%%%
where $C_{aN_i}$ denotes the ALP-RHNs Wilson coefficient and through this DM particle $N_1$ can communicate with the ALP sector. We call this ALP-portal RHN DM. We consider ALP mass of the order of few hundreds of MeV to GeV. We have considered $N_1$ to have mass up to few TeV. In the next section we discuss the allowed range for the several parameters of the model from the various phenomenological bounds.
%%%%%%%%%%%%%%%%%%%%%%%%%%%%%%%%%%%%%%%%%%%%%%%%%55
\section{Existing constraints on ALP parameter space}
\label{sec:Constraints}
%%%%%%%%%%%%%%%%%%%%%%%%%%%%%%%%%%%%
Before diving into the details of DM analyses, lets first recall the existing experimental bounds on the couplings of ALPs to gluons, photons, fermions and also from collider searches with $f_a\sim\mathcal{O}(1\,\text{TeV})$~\cite{Mimasu:2014nea,Jaeckel:2015jla,Dolan:2014ska,Izaguirre:2016dfi,Agashe:2014kda,Vinyoles:2015aba,Raffelt:2006cw,Friedland:2012hj,Ayala:2014pea,Khachatryan:2014rra,Aad:2015zva,Krnjaic:2015mbs,Clarke:2013aya,Aprile:2014eoa,Viaux:2013lha,Rodriguez-Tzompantzi:2020kuc}. ALP photon coupling is the primary parameter through which astrophysical and cosmological bounds are set on such particles. The ALP and photon coupling can be deduced
%%%%%%%%%%%%%%%%%%%%%%%%%%%%%%%%%%%%
\begin{equation}
\mathcal{L}_{a\gamma} = - C_{a\gamma} aF_{\mu\nu}\tilde{F}^{\mu\nu},\,\,\text{with}\,\,C_{a\gamma}= \frac{(C_{\tilde{B}}\cos^2\theta_w + C_{\tilde{W}}\sin^2\theta_w)}{f_a} 
\end{equation}
Particularly experiments like CAST, using primakoff process, constrained the parameter space~($M_a-C_{a\gamma}$) heavily for $M_a$ smaller then few eV~\cite{Anastassopoulos:2017ftl}. For $M_a\sim 1$~MeV the best present constraint comes from Beam Dump experiments, $C_{a\gamma}/f_a \leq 10^{-5}\,\text{GeV}^{-1}$~\cite{Mimasu:2014nea}~\footnote{For very low masses, tighter constraint exists but we do not discuss them here as we are interested in mass range $M_a\sim\text{MeV-few GeV}$.}. A slightly higher mass range is constrained by the collider experiments such as LEP and LHC~\cite{Mimasu:2014nea}. In LEP, the process like $e^-e^+\to\gamma a \to 3\gamma$ is being analysed to constrain the coupling whereas in LHC it is $pp \to \gamma a$ which searched for mono, di or tri photon signals. All these constraints are described in the references~\cite{Bauer:2018uxu,Brivio:2017ije,vinyoles2015new}. A limite on axion gauage boson coupling $C_{\tilde{W}}/f_a < 10^{-5}$ GeV$^{-1}$ is obtained for $0.175 \le M_a \le 4.78$ GeV by analysing of the process $B^{\pm} \rightarrow K^{\pm} a, a \rightarrow \gamma\gamma$ at the BABAR experiment~\cite{BaBar:2021ich}. Constraint on $C_{\tilde{G}}$ is set by mono jet 8 TeV LHC analysis. Unlike $C_{a\gamma}$, here it is much more complicated to put bound on $C_{\tilde{G}}$ due to large numbers of diagrams involved in the process. However, the dominant digram was found to be $gg\to ag$, which is complicated due to hadronisation that leads to jets in the final state. These constraints have been analysed in the reference~\cite{Mimasu:2014nea,Khachatryan:2014rra}. The limit from this study reads as $C_{\tilde{G}}/f_a \leq 10^{-4}\,\text{GeV}^{-1}$ for $M_a\leq 0.1\,\text{GeV}$. Also the bound on $\text{BR}(K^{+}\to\pi^{+}+\text{nothing})$~\cite{Adler:2004hp} can be used to constrain the process $K^{+}\to\pi^{+}+\pi^0(\pi^0\to a)$ which can be reinterpreted in terms of ALP-gluon coupling $C_{\tilde{G}}$, yielding $C_{\tilde{G}}/f_a \leq 10^{-5}\,\text{GeV}^{-1}$ for $M_a\leq 60\,\text{MeV}$. Constraints on the ALP matter coupling $C_{a\phi}$ is studied less compared to gauge boson coupling, however several processes involving flavors have been studied to put bound on it~\cite{Dolan:2014ska}. The constraints on ALP-fermion coupling $C_{a\Phi}$ depends on the ALP mass. The higher mass range is tested through the rare meson decays. Rare meson decay at Beam Dump experiments~(CHARM) sets tight constraints on $C_{a\Phi}$ for mass range $1\,\text{MeV}\leq M_a\leq 3\,\text{GeV}$ as $C_{a\Phi}/f_a<(3.4\times 10^{-8}-2.9\times 10^{-6}) \ \text{GeV}^{-1}$~\cite{Bergsma:1985qz}. For our interested mass range we summaries the existing tightest constraints in Table.~\ref{tab:constraint}. 
%%%%%%%%%%%%%%%%%%%%%%%%%%%%%%%%%%%%%%%%%%%%%%%%%%%%%
\begin{table}[h]
\begin{tabular}{|c|c|c| }
 \hline
  Bound on Coupling  & ALP Mass Range  &  Observables \\
 \hline
 $\frac{C_{\tilde{G}}}{f_a}  \leq 10^{-4}$ GeV$^{-1}$   &   $M_a \leq 0.1$~GeV  &  mono-jet 8 TeV@LHC\\
$\frac{C_{a\gamma}}{f_a}  \leq 10^{-5}$ GeV$^{-1}$   &   $M_a \sim$ 1 MeV   &  Beam Dump\\
 $\frac{C_{\tilde{W}}}{f_a} \sim 10^{-5}$ GeV$^{-1}$   &  0.175 GeV $ \le M_a \le $ 4.78 GeV   &  BABAR Exp.\\ 
 $\frac{C_{a\Phi}}{f_a} \sim  10^{-8} - 10^{-6}$ GeV$^{-1}$   &  1 MeV $< M_a <$ 3 GeV   &  Rare meson decay\\
\hline
\end{tabular}
\caption{\footnotesize{Summary of existing constraint on ALP couplings.}}
\label{tab:constraint}
\end{table}
%%%%%%%%%%%%%%%%%%%%%%%%%%%%%%%%%%%%%%%%%%%%%%%%%%%%%%%%%

\underline{\bf{Benchmark:}} To start analyzing the model, we choose ALP mass $M_a=10$ GeV. ALP-Gauge boson couplings $C_{\tilde{B}}=C_{\tilde{W}}$, since ALP is mostly constrained through $C_{a\gamma}, C_{\tilde{G}}$ and we also choose $\frac{C_{\tilde{G}}}{f_a}, \frac{C_{a\gamma}}{f_a}$ of the order of $10^{-4}$ GeV$^{-1}$ and ALP-fermion coupling $\frac{C_{a\Phi}}{f_a} \sim 10^{-6}$ GeV$^{-1}$, which satisfy all of the above constraints.  

\underline{\bf{ALP decay:}} ALP can decay into SM final states when kinematically accessible, such as leptons, gauge bosons~($W^{\pm},Z,\gamma,g$) and hadrons. Analytical form of these decay widths are listed in Appendix~\ref{app:ALP decay widths}. We found that for the range of couplings considered by us, ALP decay width is small enough to use the narrow width approximation.
%%%%%%%%%%%%%%%%%%%%%%%%%%%%%%%%%%%%%%%%%%%%%%%%%%%%%%%%%%%
\begin{figure}[h!]
\includegraphics[height=6cm, width=8cm]{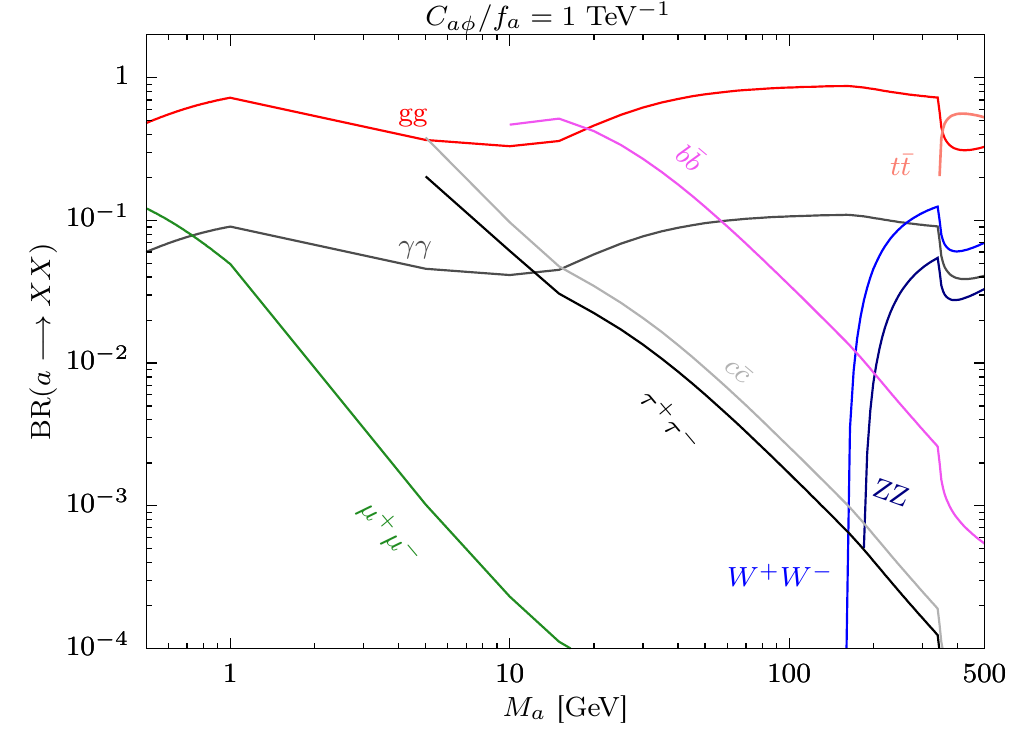}
\includegraphics[height=6cm, width=8cm]{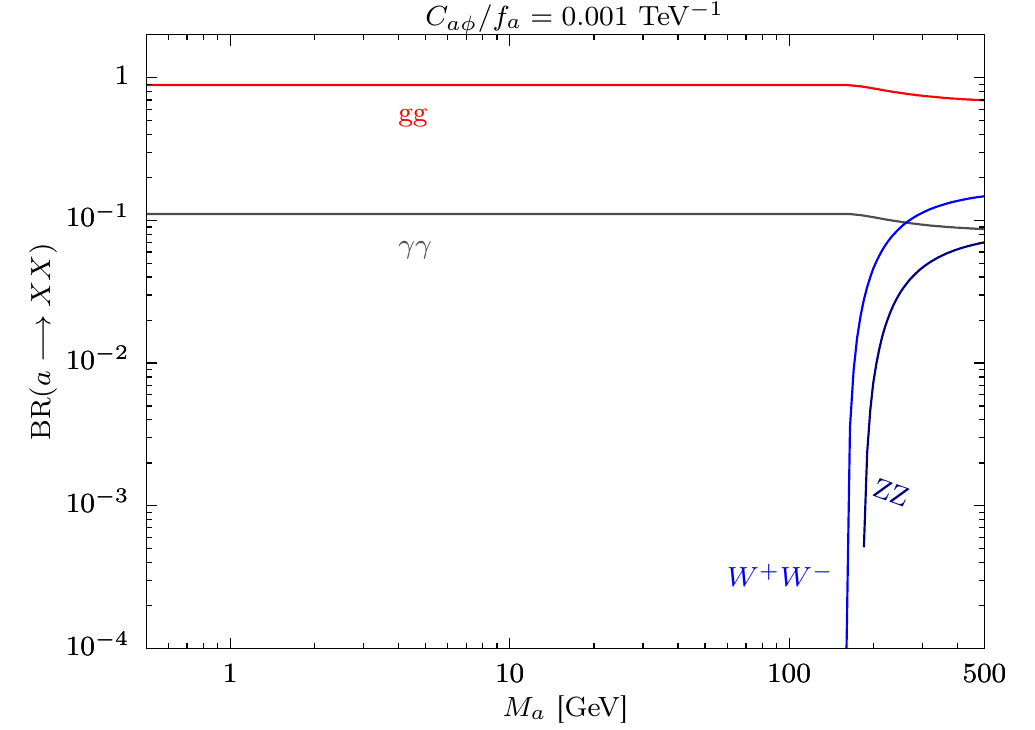}
\caption{\footnotesize{The left and right panels show the variations of branching ratios for various decay channels with respect to ALP mass $M_a$. Different colors stand for different final states. In the right panel we choose smaller value of $C_{a\phi}/f_a=10^{-3}$ TeV$^{-1}$ to show that ALP decays dominantly to gauge bosons.}}
\label{branching ratio}
\end{figure}
%%%%%%%%%%%%%%%%%%%%%%%%%%%%%%%%%%%%%%%%%%
In Fig.~\ref{branching ratio}, we show the various branching ratios of ALP decay to SM final states. For left and right panel, we choose $C_{\tilde{G}}/f_a=C_{a\gamma}/f_a=0.1\,\text{TeV}^{-1},\,C_{a\Phi}/f_a=1\,\text{TeV}^{-1}$ and $C_{\tilde{G}}/f_a=C_{a\gamma}/f_a=0.1\,\,\text{TeV}^{-1},C_{a\Phi}/f_a=0.001\,\text{TeV}^{-1}$, respectively. We see from Fig.~\ref{branching ratio} that ALP mostly decays to gluon and photon pair. Also $W^+W^-$, $ZZ$ pair contributes significantly as ALP mass crosses respective threshold values. Comparing the left panel with right panel one sees that lepton channels only contributes when $C_{a\Phi}$ is relatively large.
%%%%%%%%%%%%%%%%%%%%%%%%%%%%%%%%%%%%%%%%%%%%%%
\section{Dark matter analysis}
\label{sec:Dark matter}
%%%%%%%%%%%%%%%%%%%%%%%%%%%%%%%%%%%%%
So far we have discussed the constraints on ALP couplings to SM field. In this section we collect the results of our analyses of DM phenomenology. In our case $N_1$ plays the role of DM due to $\mathbb{Z}_2$ symmetry protection. In order to calculate all the vertices, the model is implemented in the FeynRules package~\cite{Alloul:2013bka}. All DM observables such as the thermal component of the DM relic abundance are determined using micrOMEGAS~\cite{Belanger:2018ccd} which relies on CalcHEP~\cite{Belyaev:2012qa} model file obtained from FeynRules. In the scanning of the relevant parameters of the model, we have imposed phenomenological bounds we discussed earlier. The mass of $N_{2,3}$ is always chosen to be greater than $N_1$ for further analysis.
%%%%%%%%%%%%%%%%%%%%%%%%%%%%%%%%%%%%%%%
\begin{figure}[h!]
\includegraphics[height=4cm, width=16cm]{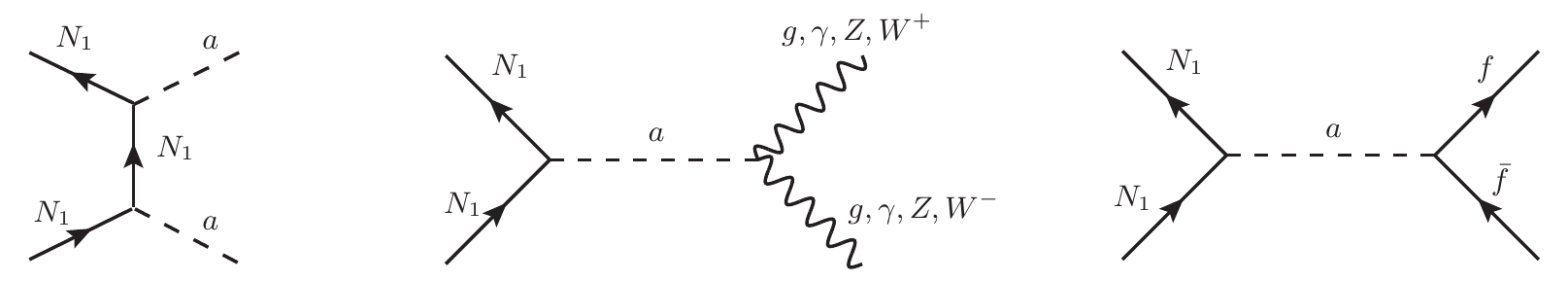}
\caption{\footnotesize{Annihilation diagrams contributing to the relic abundance of $N_1$.}}
\label{annihilation diagram}
\end{figure}
%%%%%%%%%%%%%%%%%%%%%%%%%%%%%%%%%%%%%%%
\subsection{Relic density}
\label{subsec:relic density}
%%%%%%%%%%%%%%%%%%%%%%%%%%%%%%%%%%%
The relevant processes responsible for the freeze-out of DM in the early universe are shown in Fig.~\ref{annihilation diagram}. All together, they determine the relic abundance of our assumed DM, $N_1$. The annihilation cross section$(\sigma)$ and thermal average annihilation cross section $\braket{\sigma v}$ and thus the relic abundance, scales straightforwardly with the parameters of the model. The exact analytical expressions for all the annihilation cross section are given in Appendix.~\ref{app:annihilation} and Appendix.~\ref{app:thermal average cross section}. For our considered benchmark, only gauge bosons channels are relevant. We see that annihilation cross section for the process $N_1N_1\to f\bar{f}$ is negligible due to very small value of $C_{a\Phi}$. On the other hand with the limit $M_{N_1}>M_a$, $N_1N_1\to aa$ annihilation channel opens up but is $v^2$ suppressed. 
%%%%%%%%%%%%%%%%%%%%%%%%%%%%%%%%%%%%%%%%%%%%%%%%%%%%%%%%%%%
\begin{figure}[h!]
\includegraphics[height=6cm, width=8cm]{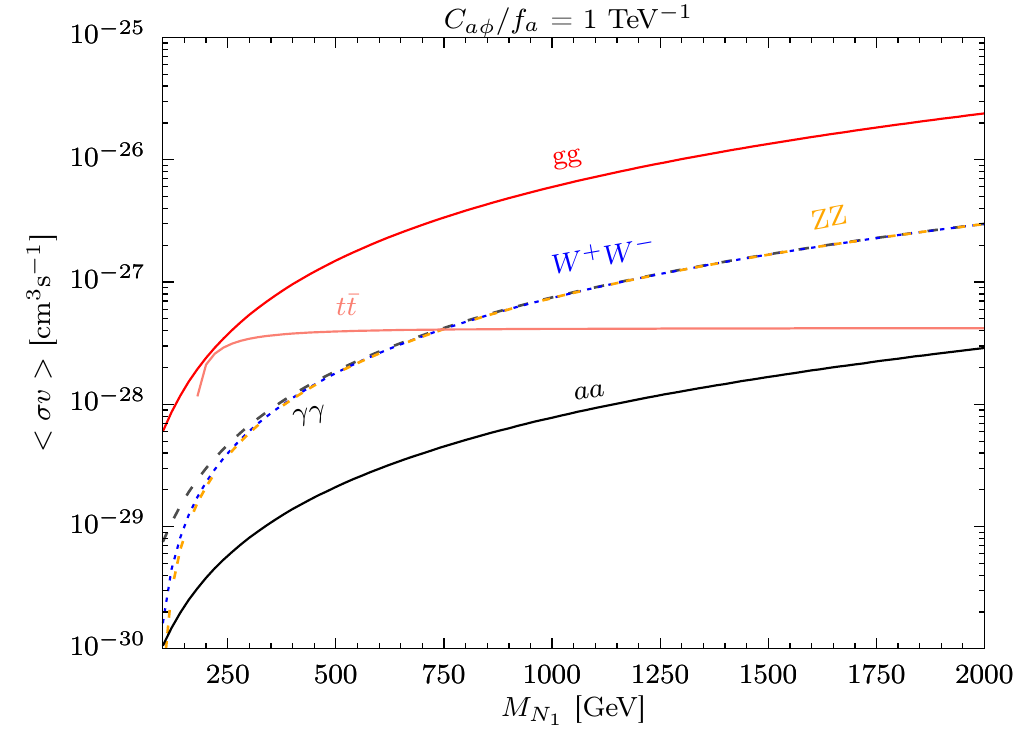}
\includegraphics[height=6cm, width=8cm]{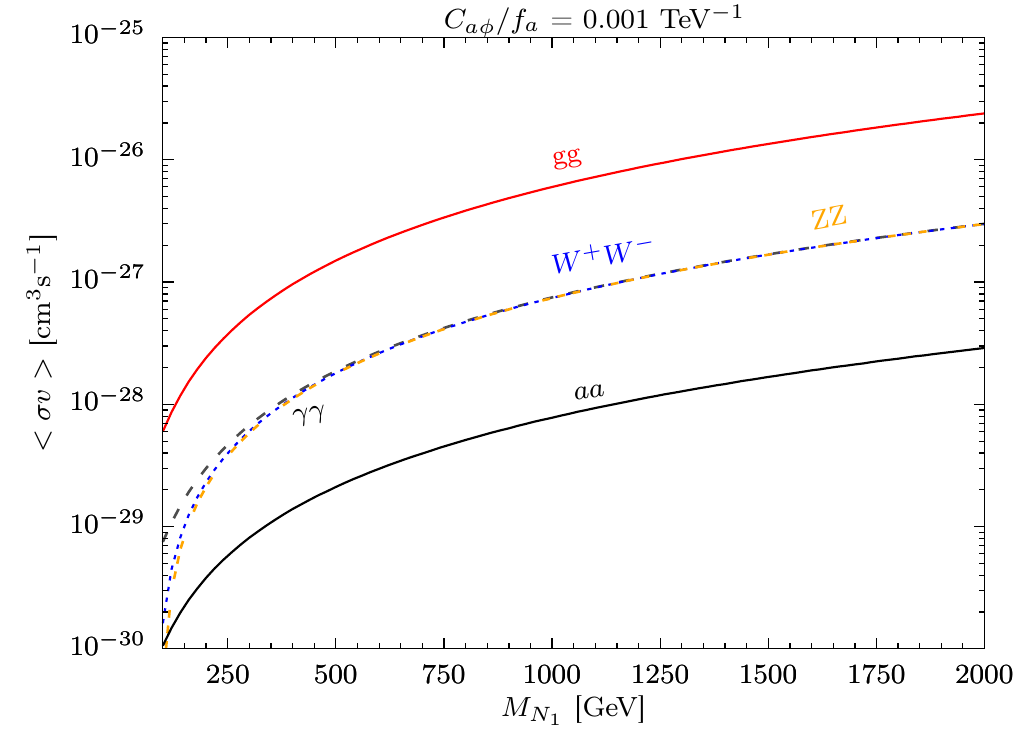}
\caption{\footnotesize{Annihilation cross sections for different channels as a function of DM mass $M_{N_1}$. In the right panel we choose smaller value of $C_{a\Phi}/f_a = 10^{-3}\,\text{TeV}^{-1}$ to illustrate that annihilation to gauge bosons dominate. In both panel we fix $C_{aN_1}=0.1\,\text{TeV}^{-1}$. }}
\label{annihilation cross section}
\end{figure} 
%%%%%%%%%%%%%%%%%%%%%%%%%%%%%%%%%%%%%%%%%
In Fig.~\ref{annihilation cross section}, we show the annihilation cross section for different channels as a function of DM mass $M_{N_1}$. For left and right panel, we choose $C_{a\Phi}/f_a=1\,\text{TeV}^{-1}$ and $C_{a\Phi}/f_a=0.001\,\text{TeV}^{-1}$ respectively. For both panel we fix $C_{\tilde{G}}/f_a=C_{a\gamma}/f_a = C_{aN_1}=0.1\,\text{TeV}^{-1}$. For relatively small value of $C_{a\Phi}/f_a$ annihilation cross section to gauge boson dominates over fermionic channels.

%%%%%%%%%%%%%%%%%%%%%%%%%%%%%%%%%%%%%%%%%%%%
\begin{figure}[h!]
\includegraphics[height=6cm, width=8cm]{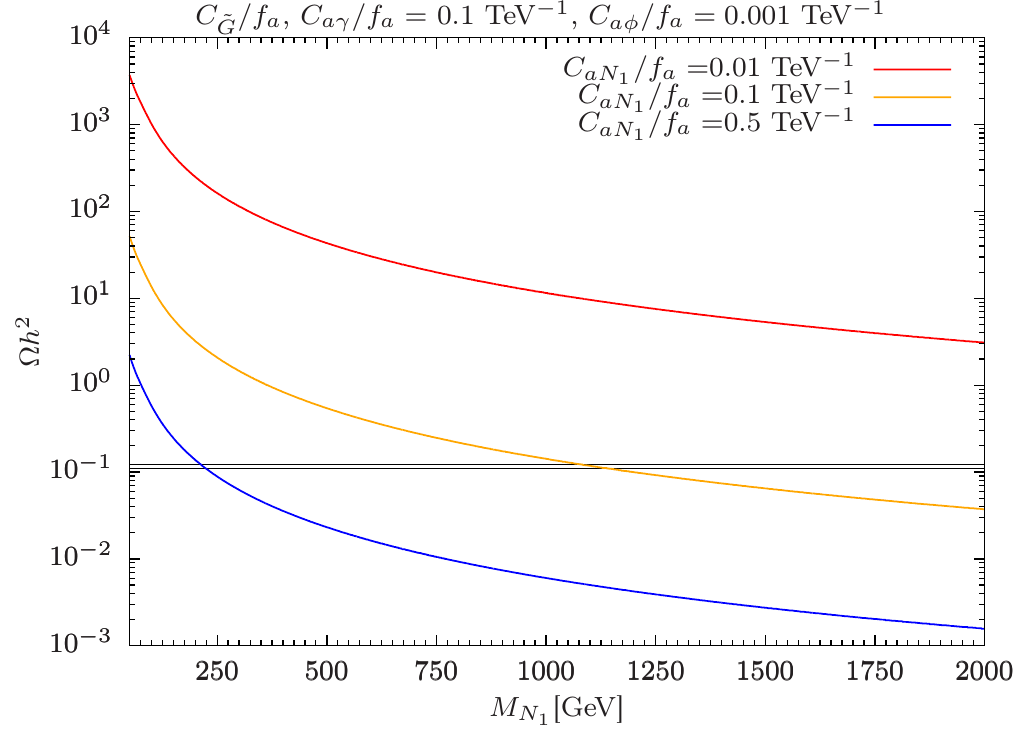}
\includegraphics[height=6cm, width=8cm]{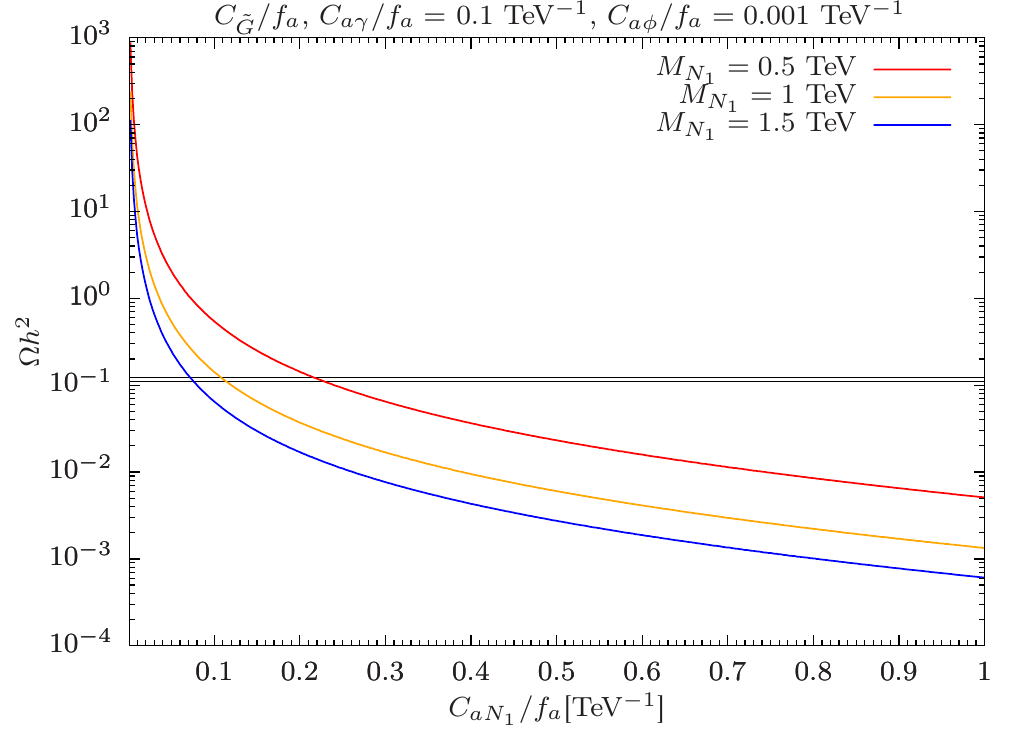}
\caption{\footnotesize{The left panel shows the relic density~($\Omega_{N_1} h^2$) behavior with DM mass $M_{N_1}$ for the choosen ALP parameters, labeled at the top. The three colored curves are due to three discrete choice of ALP-$N_1$ coupling. The region inside the horizontal black lines stands for the measured $3\sigma$ relic density range given by Planck satellite data, Eq.~\ref{eq:omega}. The right panel is done using same analysis but with ALP-$N_1$ coupling varying continuously on the horizontal axis whereas mass of $N_1$ has been chosen discretely.}}
\label{fig:relic density1}
\end{figure}
%%%%%%%%%%%%%%%%%%%%%%%%%%%%%%%%%%%%%%%%%%
The left and right panel of Fig.~\ref{fig:relic density1} shows the relic density behaviour as a function of DM mass $M_{N_1}$ and ALP-RHN coupling $C_{aN_1}$, respectively. In the left and right panel, three curves stand for three choices of ALP-RHN coupling and DM masses, respectively. The narrow horizontal band is the $3\sigma$ range for cold DM derived from the Planck satellite data~\cite{Aghanim:2018eyx}:
%%%%%%%%%%%%%%%%%%%%%%%%
\begin{align}
  \label{eq:omega}
0.1126 \leq \Omega_{N_1} h^2 \leq 0.1246.
\end{align}
%%%%%%%%%%%%%%%%%%%%%%%
Only for solutions falling exactly within this band the totality of the DM can be explained by $N_1$.  We see from Fig.~\ref{fig:relic density1} that for smaller value of DM mass $M_{N_1}$, the required values of $C_{aN_{1}}/f_a$ is relatively large to explain the correct relic density.
%%%%%%%%%%%%%%%%%%%%%%%%%%%%%%%%%%%%%%%%%%%%%
\begin{figure}[h!]
\includegraphics[height=6cm, width=8cm]{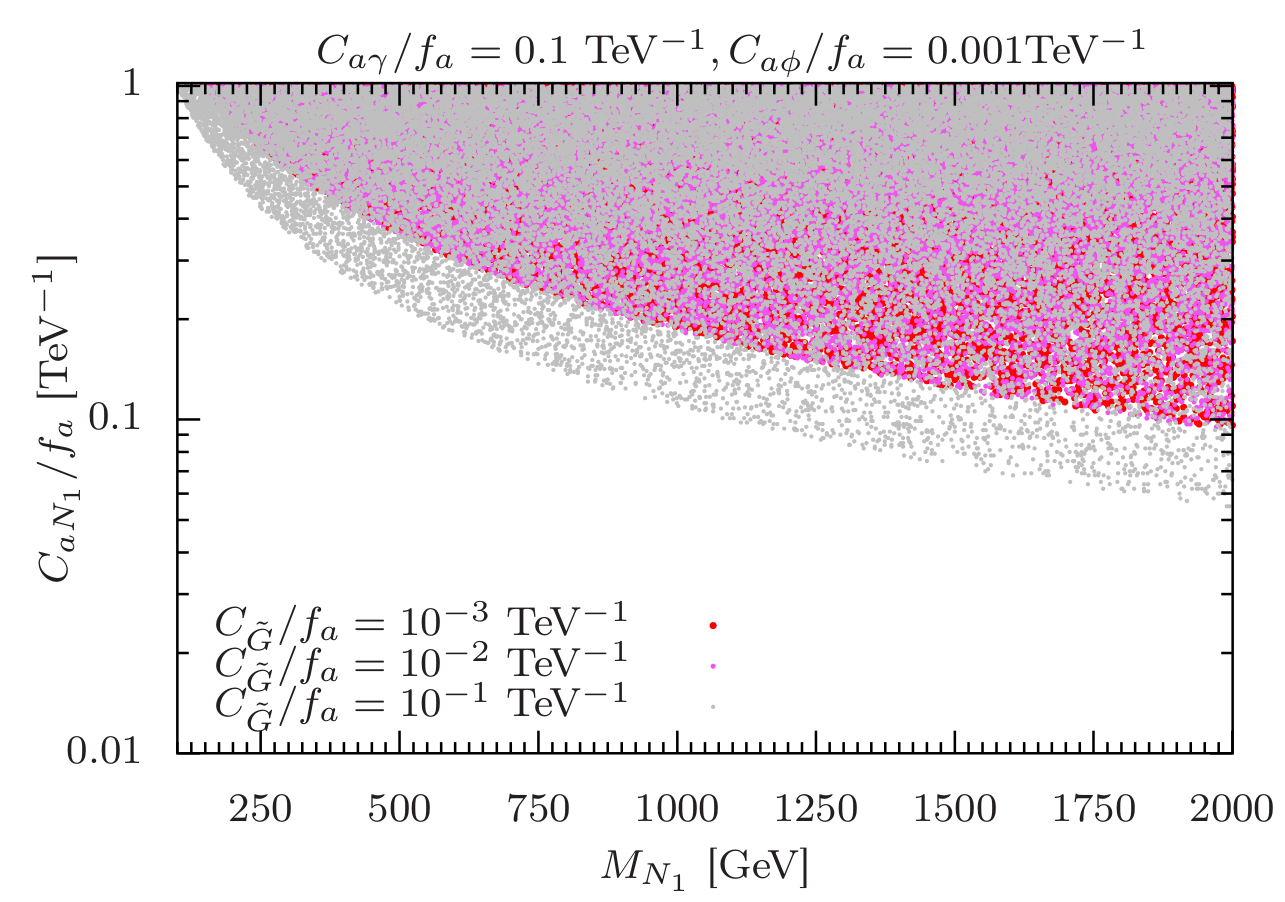}
\includegraphics[height=6cm, width=8cm]{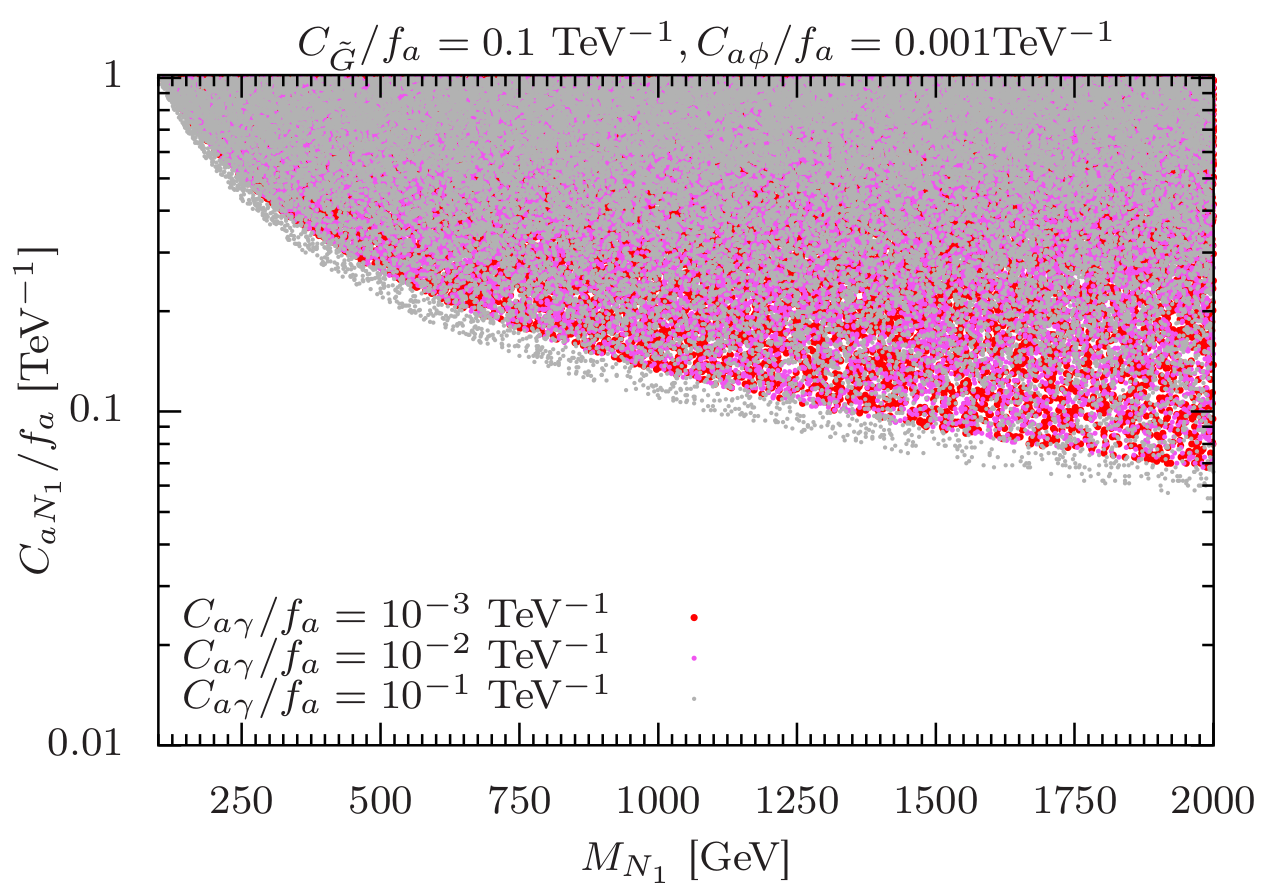}
\caption{\footnotesize{The left panel describes the colored regions where the relic density bound~($\Omega_{N_1} h^2\leq 0.12$) holds on the $m_{N_1}$-$C_{aN_1}$ plane. We have choosen the three discrete values for the ALP-gluon coupling~($C_{\tilde{G}}$) labeled by the corresponding colors. In the right panel the same analysis is done but now with three discrete values for ALP-photon coupling~($C_{a\gamma}$).}}
\label{fig:relic density2}
\end{figure}
%%%%%%%%%%%%%%%%%%%%%%%%%%%%%%%%%%%%%%%%%%%%%%%%

In Fig.~\ref{fig:relic density2} we show the region where the relic density bound $\Omega_{N_{1}}h^2\leq 0.12$ holds on the $M_{N_1}-C_{aN_1}$ plane. In the left panel we have choosen three discrete values for the ALP-gluon coupling $C_{\tilde{G}}/f_a=10^{-3}\,\text{TeV}^{-1}$~(red), $10^{-2}\,\text{TeV}^{-1}$~(pink) and $10^{-1}\,\text{TeV}^{-1}$~(gray) by fixing other couplings as $C_{a\gamma}/f_a=10^{-1}\,\text{TeV}^{-1}$ and $C_{a\Phi}/f_a=10^{-3}\,\text{TeV}^{-1}$. In the right panel, the same analysis is done but now we have fixed $C_{\tilde{G}}/f_a=10^{-1}\,\text{TeV}^{-1}$ and choose three discrete values of ALP-photon coupling, $C_{a\gamma}/f_a=10^{-3}\,\text{TeV}^{-1}$~(red), $10^{-2}\,\text{TeV}^{-1}$~(pink) and $10^{-1}\,\text{TeV}^{-1}$~(gray). Note that in left~(right) panel for relatively smaller values of ALP-gluon~(ALP-photon) couplings red and pink region overlap. This happens due to subdominant contribution to the relic density of annihilation channel $N_1N_1\to gg$~($N_1N_1\to\gamma\gamma$). We find that when $gg$ channel dominates compared to other annihilation channel, the required value of $C_{aN_1}/f_a$ is smaller to satisfy the relic density.
%%%%%%%%%%%%%%%%%%%%%%%%%%%%%%%%%%%%%%%%%%%%%%%%%
\subsection{Direct detection}
\label{subsec:direct detection}
%%%%%%%%%%%%%%%%%%%%%%%%%%%%%%%%%%%%%%%%%%%%
The XENON1T experiment~\cite{Aprile:2018dbl} currently has the best sensitivity for spin-independent and spin-dependent DM-nucleon interactions in our interested mass range of DM. The interaction between DM $N_1$ and a quark $q$ can be described by the following effective Lagrangian:
%%%%%%%%%%%%%%%%%%%%
\begin{align}
\mathcal{L}=\frac{C_{a\Phi}C_{aN_1}}{f_a^2 M_{a}^2}m_q M_{N_1} \bar{q}\gamma_5 q\,\bar{N_1}\gamma_5 N_1.
\end{align}
%%%%%%%%%%%%%%%%%%%%%%%%%%%%%%%%%%
Note that this is only valid when mediator ALP mass $M_a$ is relatively large compared to momentum transferred involved in the scattering process. Following Ref.~\cite{Boehm:2014hva,Freytsis:2010ne,Cheng:2012qr,Dolan:2014ska,Banerjee:2017wxi} we found that in non-relativistic limit, differential scattering cross section to scatter of a nucleus is $d\sigma/dE_R\propto q^4$, where $q^2=2m_N E_R$ is the momentum transfer, $m_N$ is the mass of nucleus and $E_R$ is the nuclear recoil energy. In direct detection experiments typical recoil energy is $\mathcal{O}(10\,\text{KeV})$, hence direct detection cross section is heavily suppressed.
%%%%%%%%%%%%%%%%%%%%%%%%%%%%%%%%%%%%%%%%%%%%%%%%%%%%%%%%%
\subsection{Indirect detection}
\label{subsec:indirect detection}
%%%%%%%%%%%%%%%%%%%%%%%%%%%%%%%%%%%%%%%%%%%%%%%%%%%%%%%%%%
If DM $N_1$ annihilates to SM final states with annihilation cross section near the thermal relic benchmark value $\braket{\sigma v}\sim 3\times 10^{-26}\,\text{cm}^3/s$, it may be detected indirectly. Perhaps, $\gamma$ rays are the best messengers since they proceed almost unaffected during their propagation, thus carrying both spectral and spatial information. These $\gamma$-rays can be produced from DM annihilation, either mono-energetically from direct annihilation $N_1 N_1\to\gamma\gamma$, $\gamma X$ or with continuum spectra from decays of the annihilation products $N_1N_1\to X\bar{X}(X=\text{SM state})$.  
%%%%%%%%%%%%%%%%%%%%%%%%%%%%%%%%%%
%\begin{align}
%E_{\gamma}=M_{N_1}\big(1-\frac{m_X^2}{4M_{N_1}^2}\big)\,\,\,\text{and}\,\,\,E_\gamma=M_{N_1}\,\,\text{for}\,\,X=\gamma .
%\end{align}
%%%%%%%%%%%%%%%%%%%%%%%%%%%%%
\begin{figure}[h!]
\includegraphics[height=9cm, width=12cm]{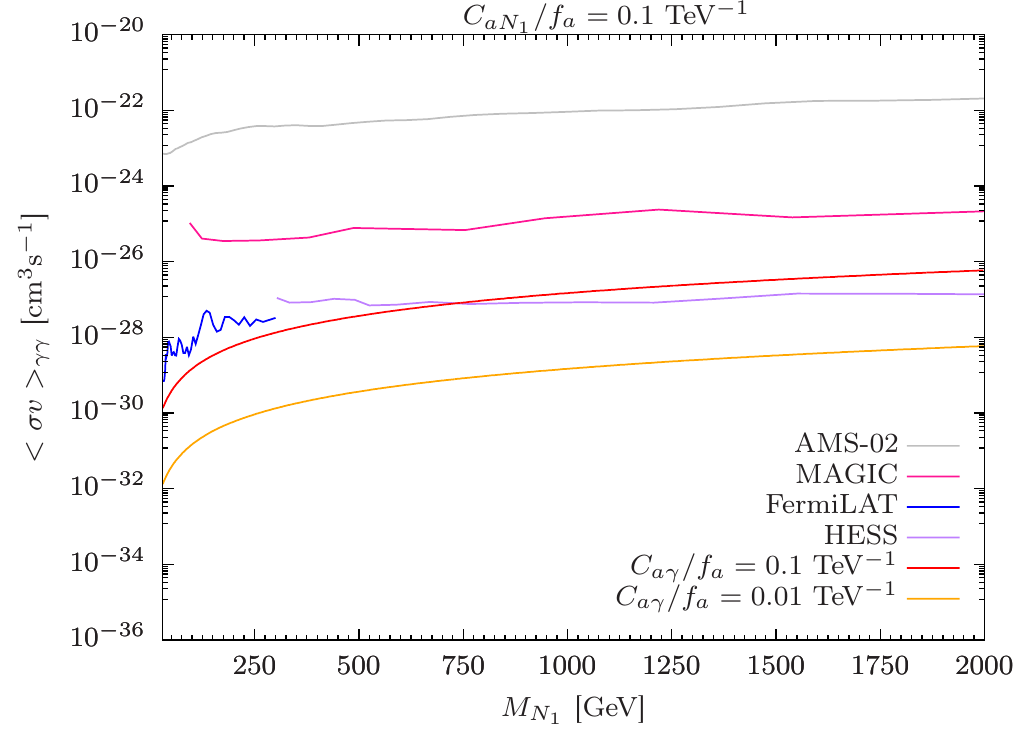}
\caption{\footnotesize{Above plot shows the thermally averaged annihilation cross section for di photon emmisions. Experimental limite obtained from AMS-02(Grey)~\cite{Elor:2015bho}, MAGIC(Dark-Pink)~\cite{HESS:2018cbt}, FermiLAT(Blue)~\cite{Ackermann:2013uma}, HESS(Purple)~\cite{HESS:2018cbt} data are shown. The red and orange lines are the the model predictions for different choices of ALP-photon couplings $C_{a\gamma}/f_a$.}}
\label{fig:indirect}
\end{figure}\\
%%%%%%%%%%%%%%%%%%%%%%%%%%%%%%%%%%%%%%%%%%%%%%%%%%%%%%%%%%
These $\gamma$ rays would be produced preferentially in regions of high DM density and can be best detectable by Fermi-LAT~\cite{Ackermann:2013uma}, HESS~\cite{HESS:2018cbt}. The integrated $\gamma$-ray flux from the DM annihilation in a density distribution $\rho(\bf r)$ is given by
%%%%%%%%%%%%%%%%%%%%%%%%%%%%%%%%%%%%%%%%%%%%%
\begin{align}
\Phi_\gamma(\Delta \Omega)=\frac{1}{4\pi}\frac{\braket{\sigma v}}{2 M_{N_1}^2}\int_{E_{\text{min}}}^{E_{\text{max}}}\frac{dN_\gamma}{dE_\gamma}dE_{\gamma}.J,
\end{align}
%%%%%%%%%%%%%%%%%%%%%%%%%%%%%%%%%%%%%%%%%%%%%%%%%%%
where $J=\int_{\Delta\Omega}{\int_{\text{l.o.s}}\rho^2(\bf r)d\ell}d\Omega'$ is the line-of-sight~(l.o.s) integral through the DM distribution integrated over a solid angle $\Delta \Omega$. The integral $\int^{\text{ROI}}\frac{dJ}{d\Omega}d\Omega$ represents the astrophysical component of the DM flux calculation in particular Region of Interest~(ROI). AMS-02 look for excess positron flux in the positron energy range $\sim$1 GeV to $\sim$500 GeV over the cosmic positron background. These positron can give rise two photon in final state after subsequent process. A model independent study of such bounds is studied\cite{Elor:2015bho}. FermiLAT collaboration look for direct DM annihilation to two photons from dwarf spheroidal galaxies of Milky Way in photon energy from few GeV to few hundreads of GeV. Higher energy range is being explored by MAGIC and HESS collaboration. We have considered the current upper limit from AMS-02, MAGIC, HESS, and annihilation data from FermiLAT respectively in Fig.~\ref{fig:indirect} on thermally averaged annihilation cross section of di-$\gamma$ final state along with our model predictions for our chosen benchmark in Fig.~\ref{fig:indirect}. The model prediction for ALP-photon coupling $C_{a\gamma}/f_a=0.1\,\text{TeV}^{-1}$ lie very close to the Fermi-LAT and HESS upper limit. This suggests that future sensitivities of Fermi-LAT or HESS can either probe or exclude large parameter space of the model considered by us.
%%%%%%%%%%%%%%%%%%%%%%%%%%%%%%%%%%%%%%%%%%%%%%%%%%%%%%%%%
\section{Conclusion}
\label{sec:Conclusion}
%%%%%%%%%%%%%%%%%%%%%%%%%%%%%%%%%%%%%%%%
We have analyzed heavy neutrino DM candidate in a minimal extension of SM, which features three RHNs and one ALP. This model is well motivated as it not only accounts for DM, but it also explains the neutrino oscillations. Hence, ALP mediated RHN DM is interesting from both the model-building and phenomenological perspectives. We have considered the lightest RHN as DM which is odd under $\mathbb{Z}_2$ symmetry and identified the region of parameters where DM predictions are in agreement with DM relic abundance.  In addition, this model also quite naturally explain the null results of LUX and XENON1T due to the pseudoscalar nature of interactions with quarks. We have highlighted the importance of complementary searches, for instance via indirect detection with single and di-photon. Although the current limits from Fermi-LAT lie above the predicted signals for our choice of parameter space, future sensitivities of Fermi-LAT might offer promising prospects to probe both the low as well as high DM mass regions.
%%%%%%%%%%%%%%%%%%%%%%%%%%%%%%%%%%%%%%%%%%%%%%%%%%%%%%%%%%%%%%%%%%%%%%%%%%%%%%%
\begin{acknowledgements}
The work of SM is supported by the Spanish grant FPA2017-85216-P (AEI/FEDER, UE) and PROMETEO/2018/165 (Generalitat Valenciana).
\end{acknowledgements}
%%%%%%%%%%%%%%%%%%%%%%%%%%%%%%%%%%%%%%%%%%%%%%%
\appendix
%%%%%%%%%%%%%%%%%%%%%%%%%%%%%%%%%
\section{ALP Decay Widths}
\label{app:ALP decay widths}
%%%%%%%%%%%%%%%%%%%%%%%%%%%%%%%
\begin{align}
&\Gamma_{agg}= \frac{2 C_{\tilde{G}} M_a^3}{\pi f_a^2},\,\,
\Gamma_{a\gamma\gamma}=\frac{M^3_a(C_{\tilde{B}}\cos^2{\theta_w} + C_{\tilde{W}}\sin^2{\theta_w})^2}{4\pi f_a^2},\,\,\Gamma_{aW^+W^-}= \frac{C_{\tilde{W}}^2(M_a^2 - 4 M_W^2)^{\frac{3}{2}}}{2\pi f_a^2} \nonumber \\
&\Gamma_{a\gamma Z}=\frac{\sin^2{\theta_w}\cos^2{\theta_w}(C_{\tilde{B}}-C_{\tilde{W}})^2(M^2_a-M^2_z)^3}{2\pi f_a^2 M^3_a},\,\,
\Gamma_{af\bar{f}}=\frac{N_c C_{a\phi}^2 M_f^2 \sqrt{M_a^2-4M_q^2}}{8\pi f_a^2} \nonumber \\
&\Gamma_{aZZ}=\frac{(C_{\tilde{W}}\cos^2{\theta_w} + C_{\tilde{B}}\sin^2{\theta_w})^2(M_a^2 - 4 M_Z^2)^{\frac{3}{2}}}{4\pi f_a^2},\,\,
\Gamma_{aN_i}=\frac{C_{aN_i}^2 m^2_{N_{i}} \sqrt{M_a^2 - 4 m_{N_{i}}^2}}{\pi f_a^2}
\end{align}
where $N_c=3(1)$ for quark(lepton).
%%%%%%%%%%%%%%%%%%%%%%%%%%%%%%%%%%%%%%%%%%%%%%%%%%%%%%%%%%%
\section{Annihilation cross sections}
\label{app:annihilation}
%%%%%%%%%%%%%%%%%%%%%%%%%%%%%%%%%%%%%%%%%%%%%%%%%%%%%%%%%
\begin{align}
&\sigma_{gg} = \frac{16 C^2_{aN_1}C^2_{\tilde{G}}M^2_{N_1}s^2}{\pi f^4_a(M^2_a-s)^2\sqrt{1-\frac{4M^2_{N_1}}{s}}},\,\,
\sigma_{\gamma\gamma} = \frac{2 C^2_{aN_1}[C_{\tilde{B}} \cos^2\theta_w + C_{\tilde{W}} \sin^2\theta_w]^2 M^2_{N_1}s^2}{\pi f^4_a(M^2_a-s)^2\sqrt{1-\frac{4M^2_{N_1}}{s}}}\nonumber \\
&\sigma_{ZZ} = \frac{2 C^2_{aN_1}[C_{\tilde{W}} \cos^2\theta_w + C_{\tilde{B}}\sin^2\theta_w]^2 M^2_{N_1}s^2(1-\frac{4M^2_Z}{s})^{3/2}}{\pi f^4_a(M^2_a-s)^2\sqrt{1-\frac{4M^2_{N_1}}{s}}},\nonumber \\
&\sigma_{W^+W^-} = \frac{2 C^2_{aN_1}C^2_{\tilde{W}} M^2_{N_1}s^2(1-\frac{4M^2_W}{s})^{3/2}}{\pi f^4_a(M^2_a-s)^2\sqrt{1-\frac{4M^2_{N_1}}{s}}},\,\,
\sigma_{f\bar{f}} = \frac{N_c C^2_{aN_1}C^2_{a\phi} M^2_{N_1}M^2_fs\sqrt{1-\frac{4M^2_f}{s}}}{2\pi f^4_a (M^2_a-s)^2\sqrt{1-\frac{4M^2_{N_1}}{s}}},\nonumber \\
&\sigma_{aa} = \frac{4C^4_{aN_1}M^2_{N_1}}{\pi f^4_a s}\sqrt{\frac{s-4M^2_a}{s-4M^2_{N_1}}}\bigg(2s-\frac{4M^4_a M^2_{N_1}}{M^4_a-4M^2_a M^2_{N_1}+ M^2_{N_1}s}\nonumber \\
&+\frac{8 M^2_{N_1}(2M^4_a - 4M^2_a s + s^2)\tanh^{-1}\left(\frac{\sqrt{(s-4M^2_a)(s-4M^2_{N_1})}}{-s+2M^2_a}\right)}{(s-2M^2_a)\sqrt{(s-4M^2_a)(s-4M^2_{N_1})}}\bigg)
\end{align}
here $\theta_w$ is the weinberg angle.
%%%%%%%%%%%%%%%%%%%%%%%%%%%%%%%%%%%%%%%%%%%%%%%%%%%%%%%%%%%%%%%%%%%%%
\section{Thermal average annihilation cross sections}
\label{app:thermal average cross section}
\begin{align}
&(\sigma v)_{gg}\approx \frac{256 C^2_{aN_1}C^2_{\tilde{G}}M^6_{N_1}}{\pi f^4_a(M^2_a-4M^2_{N_1})^2},\,\,
(\sigma v)_{\gamma\gamma}\approx \frac{32 C^2_{aN_1}[C_{\tilde{B}} \cos^2\theta_w + C_{\tilde{W}} \sin^2\theta_w]^2 M^6_{N_1}}{\pi f^4_a(M^2_a-4M^2_{N_1})^2}  \nonumber \\
&(\sigma v)_{ZZ}\approx \frac{32 C^2_{aN_1}[C_{\tilde{B}} \sin^2\theta_w + C_{\tilde{W}} \cos^2\theta_w]^2 M^3_{N_1}(M^2_{N_1}-M^2_Z)^{3/2}}{\pi f^4_a(M^2_a-4M^2_{N_1})^2}, \nonumber \\
&(\sigma v)_{W^+W^-}\approx \frac{32 C^2_{aN_1} C_{\tilde{W}}^2 M^3_{N_1}(M^2_{N_1}-M^2_W)^{3/2}}{\pi f^4_a(M^2_a-4M^2_{N_1})^2},\,
(\sigma v)_{f\bar{f}}\approx \frac{2N_c C^2_{aN_1} C^2_{a\phi} M^3_{N_1}M^2_f \sqrt{M^2_{N_1}-M^2_f}}{\pi f^4_a(M^2_a-4M^2_{N_1})^2},\nonumber \\
&(\sigma v)_{aa} \approx \frac{8C^4_{aN_1}M_{N_1}v^2\sqrt{M^2_{N_1}-M^2_a}(32 M^8_{N_1} - 64 M^2_a M^6_{N_1} + 48 M^4_a M^4_{N_1} - 16 M^6_a M^2_{N_1} + 3 M^8_a)}{3 \pi f^4_a (2M^2_{N_1}-M^2_a)^4} 
\end{align}
%%%%%%%%%%%%%%%%%%%%%%%%%%%%%%%%%%%%%%%%%%%%%%%%%%%%%
\bibliographystyle{utphys}
\bibliography{bibitem}
%%%%%%%%%%%%%%%%%%%%%%%%%%%%%%%%%%%%%%%%%%%%%%%%%%%%%%%%
\end{document}